# Unconventional magnetism in all-carbon nanofoam

A. V. Rode,[*,1] E. G. Gamaly,[1] A. G. Christy,[2] J. G. Fitz Gerald,[3] S. T. Hyde,[1] R. G. Elliman,[1] B. Luther-Davies,[1] A. I. Veinger,[4] J. Androulakis,[5] J. Giapintzakis[*,5,6]

[1]*Research School of Physical Sciences and Engineering,* [2]*Geology Department,*
[3]*Research School of Earth Sciences,*
*Australian National University, Canberra, ACT 0200, Australia*
[4]*Ioffe Physical-Technical Institute, Polytechnicheskaya 26, St. Petersburg, Russia*
[5]*Foundation for Research and Technology-Hellas, Institute of Electronic Structure and Lasers, P.O. Box 1527, Vasilika Vouton, 71110, Heraklion, Crete, Greece*
[6]*Department of Materials Science and Technology,*
*University of Crete, P.O. Box 2208, 710 03, Heraklion, Crete, Greece*

**Abstract**

**We report production of nanostructured carbon foam by a high-repetition-rate, high-power laser ablation of glassy carbon in Ar atmosphere. A combination of characterization techniques revealed that the system contains both $sp^2$ and $sp^3$ bonded carbon atoms. The material is a novel form of carbon in which graphite-like sheets fill space at very low density due to strong hyperbolic curvature, as proposed for "schwarzite". The foam exhibits ferromagnetic-like behaviour up to 90 K, with a narrow hysteresis curve and a high saturation magnetization. Such magnetic properties are very unusual for a carbon allotrope. Detailed analysis excludes impurities as the origin of the magnetic signal. We postulate that localized unpaired spins occur because of topological and bonding defects associated with the sheet curvature, and that these spins are stabilized due to the steric protection offered by the convoluted sheets.**

## 1. Introduction

Materials research and technology has been primarily based on exploring the relation between the structure and properties of well-known materials as well as processes for altering the structure and properties while having total control over the parameters, which influence the above. This type of intense activity has yielded fruitful and, in many aspects, surprising and fascinating results. Carbon is a striking example. Until relatively recently, the only known polymorphs of carbon were graphite, in which $sp^2$ hybridised carbon atoms form planar sheets in a two-layer hexagonal stacking, and diamond, in which $sp^3$ carbons form a three-dimensional framework of cubic symmetry. However, in recent decades new carbon allotropes have proliferated. A 3-layer rhombohedral stacking modification of graphite was reported in 1956[1] and lonsdaleite, the two-layer hexagonal polytype of diamond, was discovered in shocked rocks in a meteorite crater in 1967.[2] Fibrous $sp^1$-hybridised carbon has been reported from a similar environment as the mineral chaoite,[3] and synthesised in the laboratory.[4] A broad spectrum of "turbostratic graphites" are also known in which individual graphene layers show rotational and tilt disorder relative to their neighbours. A recent extreme example is provided by cores of carbon spherules in the Murchison meteorite, which are aggregates of randomly oriented single layers with additional structure defects.[5] This particular specimen is among the oldest known naturally occurring materials, since it solidified before the condensation of the Solar System. In contrast, recently synthesised materials include many more distinct forms of carbon such as fullerenes,[6,7] multi-walled carbon nanotubes[8] and single-walled carbon nanotubes,[9] which are based on a mixture of $sp^2$ and $sp^3$ hybridised carbon atoms. The structural phase space of carbon between graphite-like and diamond-like hybridisation states now has many occupants. There is even more variety in the electronic properties of various carbon allotropes, which range from superconductivity[10] to ferromagnetism[11] and tuneable electrical conductivity.[12-14] Such effects have attracted enormous attention and already been used in commercial applications.

---

[*] Corresponding authors: avr111@rsphy1.anu.edu.au; giapintz@iesl.forth.gr

A physical property of particular interest regarding all the aforementioned carbon allotropes is the magnetic susceptibility, $\chi$, since this bulk probe is related to the low energy electronic spectrum. In general, all known carbon allotropes exhibit diamagnetic susceptibility in the range of $\chi = -(10^{-5}-10^{-7})$ emu/g-Oe with the exception of: (i) polymerized $C_{60}$ prepared in a two dimensional rhombohedral phase of $\chi = +(0.25-1.3) \times 10^{-3}$ emu/g-Oe (depending on the orientation of the magnetic field relative to the polymerised planes) which shows ferromagnetism;[11] (ii) the disordered glass like magnetism observed in activated carbon fibers due to nonbonding $\pi$-electrons located at edge states;[15] and (iii) the unusual magnetic behavior observed in single wall carbon nano-horns ascribed to the Van Vleck paramagnetic contribution.[16] Although ferromagnetism in polymerized $C_{60}$ is noteworthy, the observation of a positive magnetic signal in carbon nanostructures is still a case of special interest.[17]

We have recently synthesized a new form of carbon, a cluster assembled carbon nano-foam.[18-20] The nano-foam possesses a fractal-like structure consisting of carbon clusters with an average diameter of 6-9 nm randomly interconnected into a web-like foam. This material exhibits some remarkable physical properties like the lowest measured gravimetric density (~2 mg/cm$^3$) ever reported for a solid, and a large surface area (comparable to zeolites) of 300-400 m$^2$/g.

Here we report on the equally unusual magnetic properties of the cluster-assembled nano-foam. The foam shows strong positive magnetization, some of which is lost in the first few hours after synthesis, but much of which is persistent. This paper presents a detailed study of one sample, which displayed a saturation magnetization of $M_s = +0.42$ emu/g at 1.8 K 12 months after synthesis. We assess and eliminate impurities as significant contributors to the measured properties and conclude that the observed behavior, which can neither be ascribed to paramagnetism nor to conventional soft/weak ferromagnetism, is an intrinsic property of the nano-foam.

## 2. Experimental Details

### 2.1 Synthesis

The low-density cluster-assembled carbon nanofoam was produced by high-repetition-rate (2-25 kHz) laser ablation of an ultra-pure glassy carbon target in a vacuum chamber made of stainless steel with the base vacuum ~$5\times10^{-7}$ Torr, filled with high-purity (99.995%) Ar gas, inside a 2" cylinder made of fused silica ($SiO_2$). The deposited foam was scraped off from the inner cylinder surface with plastic tweezers and placed into a glass vial for further analysis. Full details regarding the experimental conditions can be found elsewhere.[19] Here, we briefly explain some unique features of the synthesis conditions, which are very different from any previously used nano-cluster synthesis method.

The carbon vapor temperature in the laser plume, where the formation process takes place, is in the range 1-10 eV (10,000 - 100,000 K), i.e. the formation takes place in a partly ionized plasma. The high-repetition-rate laser ablation creates an almost continuous inflow of hot carbon atoms and ions with an average temperature of ~2 eV into the experimental chamber. This vapor heats the ambient gas and increases the partial density of carbon atoms in the chamber. The process of formation of carbonaceous clusters begins when the carbon density reaches the threshold density, at which the probability of collisions between carbon atoms becomes sufficiently high.

The consumption rate of carbon atoms due to the cluster formation significantly exceeds the evaporation rate by laser ablation. Thus, the formation is a non-equilibrium periodical process. We suggest that the formation process comprises periodic heating and cluster formation stages, with the time period dependent on the initial Ar density, the evaporation rate, and reaction rate, which in turn is a function of the temperature and density of the atomic carbon. It is essential that during the period of cluster formation, short in comparison to the heating period, the argon gas does not cool down, but maintains its high temperature for cluster formation. Hence, the argon (and carbon vapour) temperature from the second formation cycle onward is higher than the formation threshold temperature, which is believed to initiate $sp^2$ bonding. The still higher temperature achieved during subsequent cycles is sufficient to form $sp^3$ bonds along with the $sp^2$ bonds observed in the cluster-assembled carbon foam.

### 2.2 Characterization

In order to check the reproducibility of our results we have produced several samples, which were characterized by means of scanning electron microscopy (SEM), transmission electron microscopy (TEM), high resolution transmission electron microscopy (HRTEM), electron energy loss spectroscopy (EELS), Rutherford back-scattering (RBS), and trace elemental analysis by induction-coupled plasma mass spectrometry (ICP-MS) of acid extracts.

### 2.3 Nanostructure

Our structural studies revealed the presence of hyperbolic "schwarzite" structures.[18,20] Schwarzites are anticlastic (saddle-shaped) warped graphite-like layers, analogous to the synclastic sheets in fullerenes.[21] HRTEM images suggest periodic structures within the individual clusters, with a period of ~5.6Å[16] (see Fig.1).

## 2.4 Chemical Impurities

The possibility of magnetic contamination, during synthesis, of the nanofoam samples was thoroughly examined. The impurity content in the foam was determined from 2-MeV He$^+$ ion RBS measurements, and independently by mass spectroscopy analysis of acid extracts from the foam. Both methods show comparable low impurity contents. The RBS data indicated a total concentration of Fe-Ni of about 100 ppm atomic. For mass spectrometry, several milligrams of the foam were extracted with concentrated HCl for several hours, diluted with 2% $HNO_3$, filtered and run in a Varian Ultramass quadrupole spectrometer. Initially, three separately synthesized samples of foam were analyzed for the elements Al, Ti, V, Cr, Co, Ni, Cu, Sn, Pb. One sample, selected to be the subject of the current magnetization study, had a second portion analyzed for an extended list of elements that included Sc, Mn, Fe, Zn, Ga, In, Sb and Bi (Table 1), although we note that the molecular species ArO interferes strongly with Fe and produces a false high signal at the Fe mass numbers. The sum of all the impurity elements analyzed in the foam of this study was 415 - 465 ppm assuming 30 ppm (the Ni concentration) < Fe < 80 ppm. The dominant impurity elements in the samples were Al, Cu, Zn, Pb, Ni and probably Fe. The elemental makeup, in combination with considerable run-to-run variability of impurity content are compatible with the hypothesis that the impurities are introduced randomly as metallic dust particles (primarily aluminum, brass, solder and stainless steel) originating in the apparatus. SEM examination of one of the ablation targets did in fact reveal scattered micron-scale particles of aluminum metal, lending further support to this hypothesis.

## 2.5 Magnetization Measurements

All magnetization measurements were performed in a commercial extraction magnetometer (Maglab Exa 2000) by Oxford Instruments in the temperature range $1.8 \leq T \leq 300$ K and in applied magnetic fields up to 70 kOe. Reproducibility of the magnetic properties of the foam is demonstrated by the comparable magnetization (0.36 - 0.8 emu/g) measured in six independently synthesized batches of foam 15 days after synthesis (Table 2). The magnetization of the foam in this study (0.42 emu/g at 60 days and 12 months after the synthesis) is evidently typical for the material.

The magnetization of the sample of this study was investigated in detail for its response to temperature and applied field. Figure 2 presents the temperature dependence of the mass magnetization (closed circles) of the carbon nano-foam measured at 30 kOe in the temperature range $1.8 \leq T \leq 200$ K. The curve has been corrected for the diamagnetic background of the gelatine sample holder, which was measured independently under the same conditions (crosses in Fig. 2). The raw data of the composite (carbon nano-foam + sample holder) are represented by the open circles in Fig 2. All data were collected following zero field cooling of the sample at a heating rate of 0.3 K/min. The observed signal is positive and apparently paramagnetic (*PM*). However, the expected *PM* $1/T$-dependence was not observed, which is indicative of a non Curie-Weiss system. Hence, the magnetization isotherms at low temperatures were investigated.

Figure 3 illustrates the mass magnetization of the foam as a function of the applied magnetic field at several temperatures from 1.8K to 92K. All data have been corrected for the diamagnetic contribution of the gelatine sample holder that was again measured separately under the same conditions. The measured signal is positive and the curves show *PM*-like behaviour. Nevertheless, we observe a slight hysteresis with a well-defined coercive force at low temperatures (see inset of Fig. 3) as expected for a ferromagnet (*FM*). In Fig. 4 we present the first quadrant of the magnetization isotherm taken at 1.8K. The open circles are the magnetization data, while the solid line represents a fit of the measured magnetization values to the Brillouin function with $S = 1/2$, which corresponds only roughly to the observed behaviour.

$M(H)$ data taken at $T = 1.8K$, were plotted against $1/H$ in order to obtain the saturation magnetization of 0.42 emu/g after extrapolation to infinite field. This is equivalent to a saturation moment value of $9.0 \times 10^{-4}$ $\mu_B$ per carbon atom ($\mu_B$ is the Bohr magneton equal to $9.27 \times 10^{-21}$ erg/G). Assuming that our spin system is *FM*-like rather than *PM* (i.e., 1 $\mu_B$ per unpaired spin), we estimate that this value corresponds to about 1 unpaired spin per 1000 carbon atoms, a fact which suggests that several unpaired spins are located in each of the nanometre-scale spheroidal clusters with $\sim 10^4$ C-atoms/cluster that constitute the foam,[18] and is in order-of-magnitude agreement with low temperature ESR measurements that show a large concentration of unpaired spins $1.8 \times 10^{20}$/g (3.6 unpaired spins per 1000 carbon atoms).

### 3. Discussion of Magnetization Data

We have observed a strong positive magnetization signal in a new all-carbon structure, which seems to have features of both a *PM* and a *FM*. It could be hypothesized that our observed sample behaviour combines a *FM* signal from chemical impurities (the 3*d* elements) and *PM* from the new carbon phase (foam). However, the experimental evidence does not support this possibility. We now show that the observed magnetic behaviour cannot arise from ferromagnetic impurities but rather is an intrinsic property of the nano-foam.

First, consider the magnitude of the observed signal. Both the hysteretic behaviour as well as the relative failure of the Brillouin model might be attributable to the existence of *FM* impurities (see Table 1). Non-lanthanide *FM* compounds have saturation magnetizations corresponding to effective Bohr magneton numbers of 0.5 - 3.5 $\mu_B$ per magnetic atom.[22] If we assume an absolute worst-case scenario in which all the Fe, Co and Ni were in their ferromagnetic elemental forms, the impurity magnetization would be at most 0.09 emu/g in our sample, which is only 20% of the measured total. We note that even Fe-Ni are likely to be present as non-*FM* steel. The 3*d* elements Sc-Cu make up only half of the impurity total, and the elements Fe-Ni are less than half again of that fraction. Many of the impurity elements are most likely to be present in non-ferromagnetic phases such as Al metal, brass and solder.

Furthermore, the observed response of sample magnetism to temperature and applied magnetic field is not what would be expected from transition metal bearing ferromagnetic impurities. A ferromagnetic contribution from impurity elements should remain constant (saturated) at a specific temperature for sufficiently high magnetic field values (a few kOe for normal *FM* elements) and should also change only slightly with temperature given the fact that normal 3*d*-FM elements exhibit a critical Curie temperature $T_c$ of the order of ~1000K ($T_c$ = 1043K, 1388K and 627K for Fe, Co and Ni metals respectively), as do many of their intermetallic compounds with B-group semimetals (eg MnBi, $T_c$ = 630K; $Cu_2MnAl$, $T_c$ = 710K).[22]

Because of the relative temperature insensitivity, we would expect an "impurity" signal to persist at high temperatures. However, the magnetism of our sample does not behave in that way, as is evident from Fig. 5, where we have plotted the total measured signal of the composite (foam + gelatine holder) at $T$ = 5K and $T$ = 110K as well as the measured signal of the sample holder alone as a function of the applied magnetic field. If the total measured signal was predominantly due to *FM* impurities, then the same pattern should have been observed at quite elevated temperatures. Nevertheless, above ~100K the remaining signal from the composite is comprised almost entirely of the nearly temperature-independent diamagnetic signal of the gelatine sample holder. The open circles represent a *M vs H* measurement at $T$ = 5K of the gelatine sample holder and is plotted for clarity on the same graph (Fig. 5). The strong applied field dependence of the total signal is clear. These new data indicate that the observed magnetic behaviour is extremely unlikely to arise in ferromagnetic impurities.

We now turn to discuss the "*PM*"-like signal of Figs 3 and 4. It is well known that the magnetization of a *PM*, when plotted against normalized *H/T* for several different temperatures, should collapse onto a single curve. Figure 6 shows a plot of the magnetization of the foam versus *H/T*. Data for different *T* do not scale similarly, a clear indication that the sample is not simply paramagnetic (or super-paramagnetic). Although, the "*FM*-impurities" argument could account for such an effect, this conclusion is contra-indicated by the observation that the magnetization is higher at higher temperatures, i.e. the *M(H/T)* curve at 5K is above that of the curve extracted at 3K. The same is true for the curve at 3K and so on. If the magnetization were dominated by "impurity" *FM* interactions then the situation would be reversed. A similar effect to that in our sample has recently been observed in amorphous magnetic rare earth silicon alloys[23] where the formation of ferromagnetic polarons where found to play a crucial role. The authors of Ref. 23 have speculated that competition of intra-polaron *FM* interactions and inter-polaron antiferromagnetic interactions due to significant polaron overlapping can explain their data.

Another possible source of positive susceptibility is molecular oxygen. Molecular oxygen – one of the most abundant non-metallic paramagnets in nature – is a possible source of contamination since it can potentially be trapped as an adsorbate on a high surface area material such as our foam. However, intercalation into the 5-6 Å gaps between sheets is unlikely, since the sum of two carbon Van der Waals radii (1.7 Å each) and two oxygen radii (1.5 Å each) is 6.4 Å. We note that the data reported in the current study were collected from samples that were handled in pure helium atmosphere in a glove box. Nevertheless, it was deemed prudent to perform the following independent experiment to determine an upper bound on the contribution of oxygen to our measurement. We filled a gelatin capsule, same as used in our measurements, with pure oxygen gas and then measured (i) the magnetic moment vs. magnetic field at $T$ = 5 K and (ii) the magnetic moment vs. temperature at $H$ = 3 Tesla. We observed a negative total magnetic moment due to the diamagnetism of the capsule, exactly as observed for our background measurements. Any paramagnetic contribution of oxygen was too small to be observed and hence cannot have significantly perturbed our data.

Finally, we note another experimental observation that is incompatible with conventional *FM* impurities. The produced carbon nano-foam was found to exhibit a strong magnetic relaxation in time, i.e. the measured magnetic moment decreased over a period of a few days. Surprisingly, immediately after production the foam is attracted to a permanent 4 kOe $Nd_2Fe_{14}B$ magnet at room temperature, thus demonstrating the existence of a substantial permanent magnetic moment. However, the magnetization decreases sufficiently fast so that the magnet attraction effect can no longer be observed a few hours after synthesis. The filled circles in Fig. 7 shows the magnetization of the nano-foam 15 days after production and the open circles is the same measurement 60 days after production. As evident from the above, the magnetization exhibited by the sample relaxes to lower values as a function of time, achieving magnetic equilibrium several weeks after production. We continued measuring the same sample at various intervals over the next 12 months, without any additional relaxation becoming apparent. Therefore, Figs 2-6 represent the equilibrated magnetization values.

## 4. Conclusions

The magnetic behaviour exhibited by the new phase of carbon is extremely unusual. It differs from the weak positive magnetization found at very low temperatures in single-walled nanohorns and activated carbon fibers, mentioned in the introduction. In these two cases, the occurrence of magnetism has been associated with exposed graphitic edges. Our observations point to a different, unique origin for the magnetism in the foam. We believe that the remarkable magnetic properties of the foam, unexpected for an all-carbon material, are an intrinsic consequence of its equally remarkable nanostructure.

HRTEM shows convoluted graphite-like layers inside the nano-spherular building blocks of the foam, the contrast being consistent with hyperbolic "schwartzite" curvature of the sheets.[18] This requires rings of 7 or more carbons interspersed with normal graphitic 6-rings.[24] The sheet curvature localizes unpaired spins by breaking the continuity of the delocalized π-electron clouds of graphite, and tight curvature of the sheets provides a mechanism for sterically protecting the unpaired spins which would otherwise be too chemically reactive to persist.[20]

A possible mechanism for magnetic moment generation would be a simple indirect exchange interaction through conduction electrons located on the hexagons. This, however, has to be in agreement with electrical resistivity measurements that show a semiconducting behaviour with a band gap of 0.5-0.7 eV.[18] A plausible scenario would be that the magnetism of our nano-foam is an effect that actually occurs in nano-sized metallic ($sp^2$) segments of the structure that are isolated by non-conducting ($sp^3$) regions, and hence do not contribute to the overall conductivity of the sample.

In summary, we have observed unique magnetic behaviour in an all-carbon nano-structure, the unusual structure of which provides a plausible mechanism for generation of strong magnetism. Our data leads us to reject ferromagnetic impurities as the origin of the observed magnetism. Combining our experimental results, *M vs H/T* scaling, magnitude and temperature dependence of the moment of ferromagnetic impurities, and strong time-dependent magnetization relaxation, lead us to safely conclude that the observed behaviour is an intrinsic property of the foam itself. The behaviour does not fit into the categories of conventional paramagnetism or superparamagnetism (data do not fit the Brillouin function with $S = 1/2$, and $M$ vs $H/T$ curves do not collapse onto one another). We have observed small hysteresis and remnant magnetization in the $M(H)$ curve of our foam which are usually observed in organic ferromagnets[25] and hence do not exclude the case of weak soft ferromagnetism. Nevertheless, we have no clear signs of an ordering temperature and most importantly the structure of our foam suggests that the clusters of the foam are too small to sustain a permanent magnetic moment as in normal ferromagnetic materials.

This new form of carbon clearly warrants further theoretical and experimental investigations.

## Acknowledgments

The authors are grateful to Prof. D. Tomanek of Michigan State University, for many encouraging discussions.

## References

1. F. Laves and Y. Baskin, *Zeitschrift für Kristallographie* **107**, 337-356 (1956).
2. C. Frondel and U. B. Martin, *Nature* **214**, 587-589 (1967).
3. A. El Goresy and G. Donnay, *Science* **161**, 363-364 (1968).
4. Yu. P. Kudryatsev, R. B. Heimann and S. E. Evsyukov, *J. Materials Science* **31**, 5557-5571 (1996).
5. P. Fraundorf and M. Wackenhut, *Astrophysical Journal* **578**, L153-L156 (2002).
6. H.W. Kroto, J.R. Heath, S.C. O'Brien, R.F. Curl and R.E. Smalley, *Nature* **318**, 162-3, (1985).
7. W. Krätschmer, L.D. Lamb, K. Fostiropoulos, D.R. Huffmann, *Nature* **347**, 354-358, (1990).
8. Iijima S., *Nature* **354,** 56 (1991).
9. Iijima S. and Ichihashi T., *Nature* **363**, 603 (1993).
10. C. N. R. Rao, R. Seshadri, A. Govindaraj and R. Sen, *Materials Science and Engineering: R: Reports* **15**, 209 (1995).
11. T. L. Makarova, B. Sundqvist, R. Höhne, P. Esquinazi, Y. Kopelevich, P. Scharff, V. A. Davydov, L. S. Kashevarova, A. V. Rakhmanina, *Nature* **413**, 716-718 (2001).
12. J-C. Charlier, P. Lambin, T. W. Ebbesen, *Phys. Rev. B*, **54**, R8377-R8380 (1996).
13. R. Saito, M. Fujita, G. Dresselhaus, M. S. Dresselhaus, *Phys. Rev. B*, **46**, 1804-1811 (1992).
14. L. C. Venema, V. Meunier, Ph. Lambin, C. Dekker, *Phys. Rev. B*, **61**, 2991-2995 (2000).

15. Yoshiyuki Shibayama, Hirohiko Sato, and Toshiaki Enoki, *Phys. Rev. Lett.,* **84**, 1744 (2000).
16. S. Bandow, F. Kokai, K. Takahashi, M. Yudasaka, S. Iijima, *Appl. Phys.* A, **73**, 281 (2001).
17. Lei Liu, G. Y. Guo, C. S. Jayanthi, and S. Y. Wu, *Phys. Rev. Lett.* **88**, 217206 (2002).
18. A. V. Rode, S. T. Hyde, E. G. Gamaly, R. G. Elliman, D. R. McKenzie, S. Bulcock, *Appl. Phys*. **A69**, S755-S758 (1999).
19. A. V. Rode, E. G. Gamaly, B. Luther-Davies, *Appl. Phys.* **A70**, 135-144 (2000).
20. A. V. Rode, R. G. Elliman, E. G. Gamaly, A. I. Veinger, A. G. Christy, S. T. Hyde, B. Luther-Davies, *Appl. Surf. Sci.,* **197-198**, 644-649 (2002).
21. D. Vanderbilt and J. Tersoff, *Phys Rev Lett.,* **68** 511-513 (1991).
22. C. Kittel. Introduction to Solid State Physics (4$^{th}$ edition). John Wiley and Sons, New York, London, Sydney, Toronto (1971).
23. F. Hellman, M. Q. Tran, A. E. Gebala, E. M. Wilcox, and R. C. Dynes, *Phys. Rev. Lett.* **77**, 4652 (1996).
24. S.T. Hyde and M. O'Keeffe, *Phil. Trans. R. Soc. Lond.* **A 354**, 1999 (1996).
25. A. Mrzel, A. Omerzu, P. Umek, D. Mihailovic, Z. Jaglicic, Z. Trontelj, *Chem. Phys. Lett.* **298**, 329 (1998).

**Table 1**: Trace element analysis (ICP-MS) of carbon foam samples. Fe determination is a maximum value due to $^{40}Ar^{17}O$ interference at same mass number. Stated uncertainties are standard deviations for 5 replicate measurements as an indication of spectrometer reproducibility.

| Element | Atomic ppm |
|---|---|
| Al | 125 ±4 |
| Sc | 0.5 ±0.3 |
| Ti | 0.4 ± 7 |
| V | 0.2 ±4 |
| Cr | 11 ±1 |
| Mn | 5.5 ±0.3 |
| Fe | <80 |
| Co | 0.65 ±0.07 |
| Ni | 29.7 ±1.5 |
| Cu | 110 ±7 |
| Zn | 67 ±5 |
| Ga | 0.4 ±0.2 |
| In | 0.12 ±0.02 |
| Sn | 5.6 ±0.5 |
| Sb | 0.19 ±0.09 |
| Pb | 24 ±1 |
| Bi | 3.8 ±0.2 |

**Table 2**: Magnetization measurements of carbon foam samples 15 days after synthesis. Magnetization measured at **T = 5K, H = 60 kOe.**

| Sample | Magnetic Condition | Mass Magnetization (emu/g) |
|---|---|---|
| 040602 | Paramagnetic | 0.446 |
| 050602#1 | Paramagnetic | 0.366 |
| 050602#2 | Paramagnetic | 0.8 |
| 060602#1 | Paramagnetic | 0.375 |
| 060602#2 | Paramagnetic | 0.78 |
| 070602 | Paramagnetic | 0.5 |

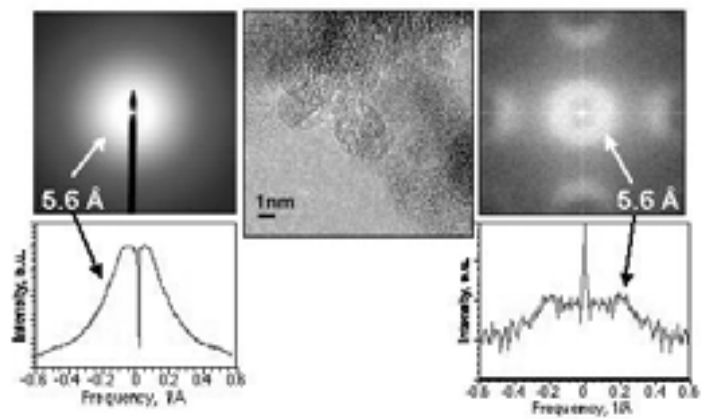

Fig. 1. High-resolution TEM image of the carbon nanofoam foam (central image) with an individual 6-nm cluster clearly seen in the centre. Fourier transforms (right) and electron diffraction patterns (left) indicate a 'repeat' spacing with characteristic length of 5.6±0.4 Å typical for 'schwarzites'.

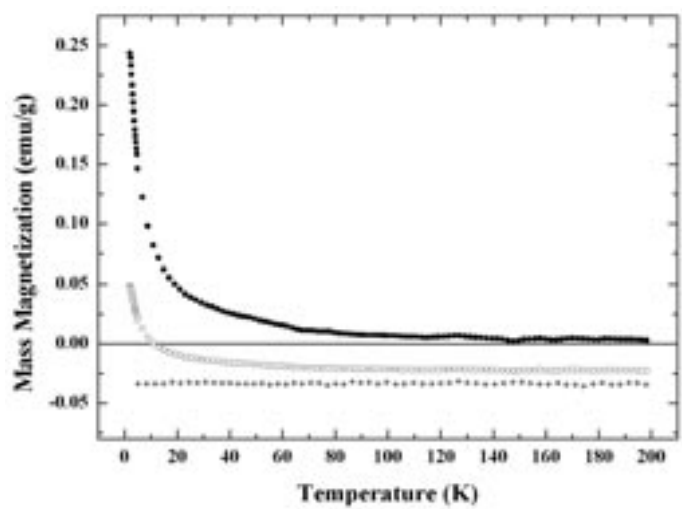

Fig. 2. Temperature dependence of the mass magnetization of the composite of carbon nano-foam and sample holder (open circles), the carbon nano-foam (filled circles) and the gelatine sample holder (crosses) measured at 30 kOe in the temperature range $1.8 \leq T \leq 200$ K.

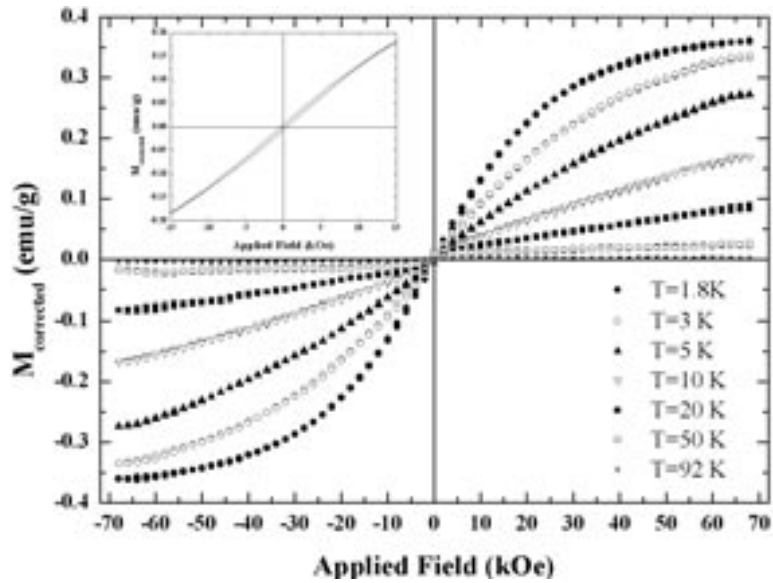

Fig. 3. Mass magnetization of the foam as a function of the applied magnetic field, *M(H)*, at several temperatures from 1.8 to 92 K. All data are corrected for the diamagnetic contribution of the gelatine sample holder. Inset: *M(H)* hysteresis loop at T = 1.8 K exhibiting a coercive force $H_c$ = 420 Oe.

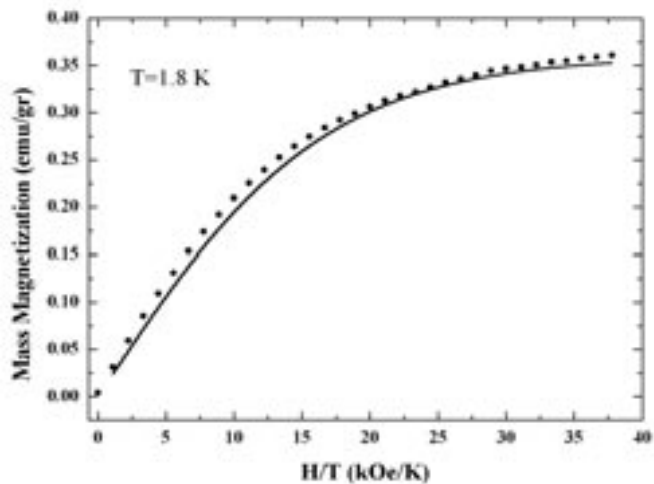

Fig. 4. Corrected mass magnetization of the foam as a function of H/T measured at T =1.8 K. Solid line is a fit of the data to the Brillouin function with S =1/2.

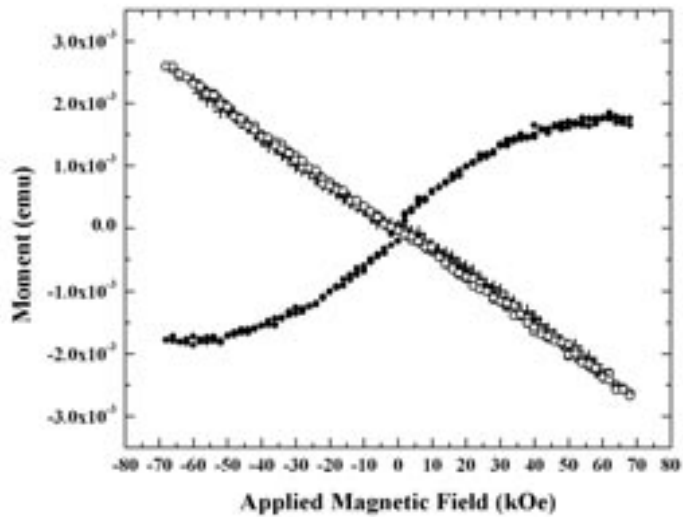

Fig. 5. Magnetic moment as a function of the applied magnetic field for the gelatin sample holder measured at $T = 5$ K (crosses), and the composite of carbon nano-foam and sample holder measured at $T = 5$ K (closed circles) and $T = 110$ K (open circles).

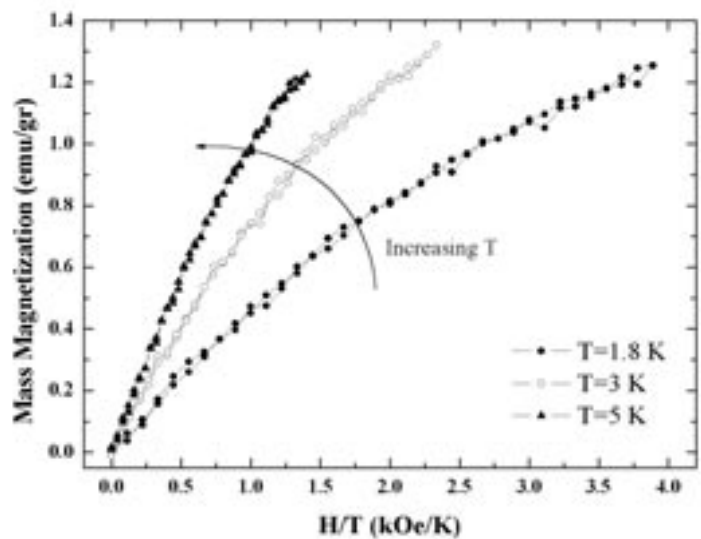

Fig. 6. Corrected mass magnetization as a function of $H/T$ at $T = 1.8$ K (closed circles), 3 K (open circles) and 5 K (triangles). Note that the curves do not scale as it is expected for a *PM* and have the opposite behaviour from that expected for a *FM*.

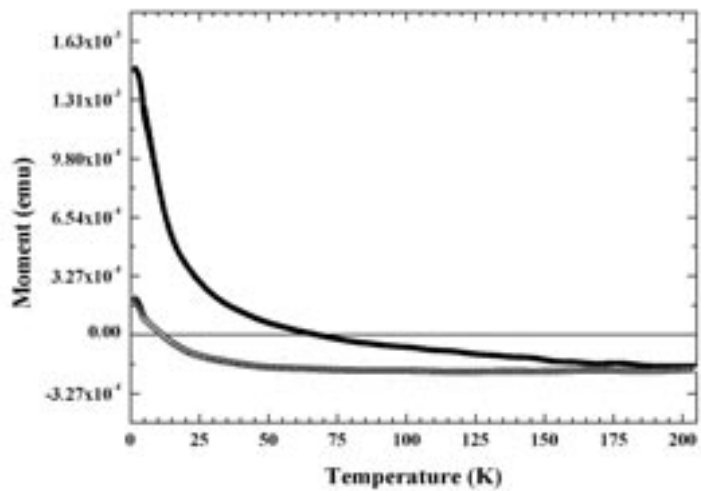

Fig. 7. Temperature dependence of the magnetic moment of the composite of carbon nano-foam and sample holder measured at *H* = 30 kOe, 15 days (filled circles) and 60 days (open circles) after production of the sample.